# Mobility Extraction and Analysis of GaN HEMTs for RF Applications Using TCAD and Experimental Data

Tanjim Rahman

Department of Electrical & Computer Engineering

Texas Tech University, Texas, USA

tanjrahm@ttu.edu

*Abstract*— **This paper presents an analysis of GaN high-electron-mobility transistors (HEMTs) using both TCAD simulation and experimental characterization. The energy band structure was studied using Nextnano simulation software to observe two-dimensional electron gas (2DEG) formation and carrier confinement under equilibrium conditions. Additionally, I-V and C-V data from fabricated research-grade GaN HEMTs were analyzed to extract key electrical parameters. The device demonstrated an ON current of 1.9 mA and an OFF current of 0.01 mA, indicating a strong ON/OFF current ratio. A subthreshold swing of 80 mV/decade and a DIBL of 5 mV/V were observed, confirming good gate control and short-channel suppression. The ON-resistance was 22.72 Ω·µm, with a saturation voltage of 1 V. The peak transconductance was extracted as 0.18 mS in the linear region and 0.5 mS in saturation. Field-effect mobility was calculated using the transconductance method, with a maximum value of approximately 1200 cm²/V·s at low drain bias. The combined simulation and experimental approach provided comprehensive insight into GaN HEMT behavior, enabling a deeper understanding of structure-performance relationships critical to advanced transistor design.**

*Index Terms*—GaN HEMT, ON/OFF current ratio, subthreshold swing, DIBL, ON-resistance, transconductance, power electronics, RF applications.

I. INTRODUCTION

GaN HEMTs are among the most promising wide bandgap semiconductor devices, offering exceptional performance for high-power, high-efficiency, and high-frequency electronic applications across a variety of advanced technologies and emerging systems[1]. Gallium nitride (GaN) based high-electron-mobility transistors (HEMTs) have become increasingly important in modern electronics, especially for high-frequency and high-power applications[2]. The GaN high electron mobility transistors (HEMTs) have a high breakdown voltage with high cut-off frequency, compared to the other material based devices, leading to high power systems with high efficiency[4-7]. The formation of a two-dimensional electron gas (2DEG) at the AlGaN/GaN heterointerface, caused by conduction band discontinuity, significantly enhances carrier transport in HEMTs [8]. The presence of 2DEG eliminates the need for doping in the channel, allowing for faster switching and lower conduction losses[9].

This paper addresses both theoretical and experimental analyses of GaN HEMT operation. Equilibrium energy band diagrams were simulated using Nextnano to elucidate 2DEG formation and carrier confinement mechanisms in the heterostructure. Complementary I-V and C-V measurements from fabricated devices enabled extraction of key performance parameters, including threshold voltage, ON/OFF current ratio, subthreshold swing, DIBL, ON-resistance, and transconductance. The combined simulation-measurement approach provides comprehensive and in-depth insight into the influence of heterostructure design on electrical behavior and overall device performance, guiding the optimization of GaN HEMT devices for enhanced efficiency, scalability, and high-frequency operation in advanced electronic systems.

In this paper Section II presents the band structure analysis using Nextnano simulation results, Section III covers experimental measurements, parameter extraction and corresponding discussion, and Section IV concludes the paper with key observations and implications.

## II. STRUCTURE ANALYSIS THROUGH NEXTNANO FOR BAND DIAGRAM

To study the band structure and carrier behavior in GaN/AlGaN HEMTs, simulations were conducted using Nextnanos software. The modeled structure included a GaN channel and an AlGaN barrier with realistic layer thicknesses, mole fractions, and doping profiles to closely match fabricated device configurations. The conduction band diagram under equilibrium conditions clearly showed the formation of a two-dimensional electron gas (2DEG) at the AlGaN/GaN interface, driven by strong polarization-induced electric fields. This 2DEG plays a crucial role in enabling high electron mobility and current drive capabilities. The simulation also demonstrated how variations in barrier thickness, Al mole fraction, and donor concentration significantly influence band bending, carrier confinement, and the resulting sheet charge density. Such insights are essential in understanding how electrostatic design impacts device operation. These results highlight the critical role of heterostructure engineering and precise material control in enabling efficient electron transport and optimizing HEMT performance for high-frequency and high-power applications.

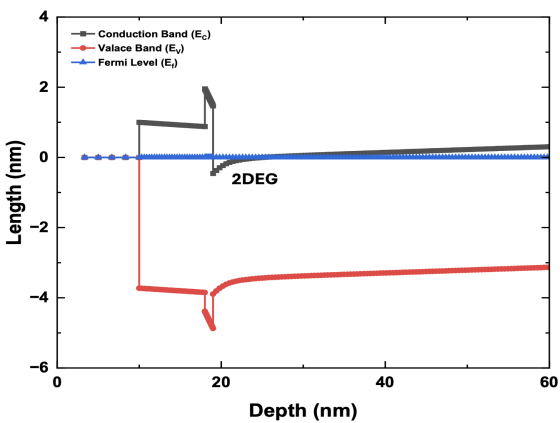

Fig. 1. Nextnano-simulated energy band diagram of a GaN/AlGaN HEMT under equilibrium conditions.

## III. DEVICE MEASUREMENT & RESULT ANALYSIS

The AlGaN/GaN HEMT was electrically characterized under standard laboratory conditions using a probe station integrated with a Keysight B1500A Semiconductor Parameter Analyzer. Both current–voltage (I-V) and capacitance–voltage (C-V) measurements were performed to extract key device parameters.

C–V measurements were conducted at a frequency of 1 MHz to evaluate the gate capacitance characteristics. For transfer characteristics ($I_D$–$V_g$), measurements were performed in both the linear and saturation regions. In the linear region, the drain-source voltage ($V_{DS}$) was swept from 10 mV to 100 mV in 10 mV increments, ensuring the device operated in the ohmic region for accurate transconductance analysis. In the saturation region, $V_{DS}$ was swept from 100 mV to 1 V in 100 mV steps to capture current behavior under high-field conditions.

Output characteristics ($I_D$–$V_d$) were measured by sweeping $V_{DS}$ while setting $V_{GS}$ from –2 V to 0 V in 0.2V increments. These settings enabled the extraction of output resistance, saturation behavior, and knee voltage, which are critical for evaluating the power handling and switching capability of the device.

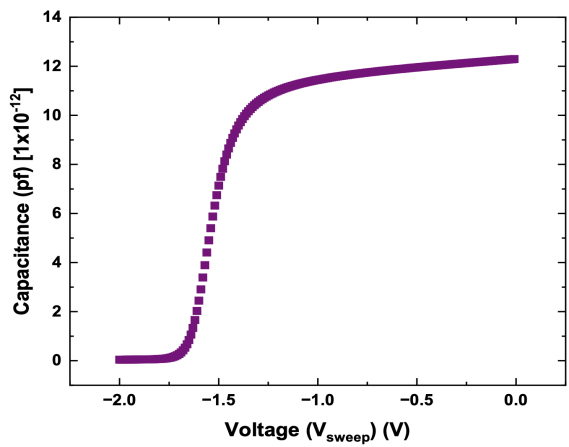

Fig. 2. Measured C–V characteristics of AlGaN/GaN HEMT at 1 MHz.

Fig. 2 shows the C–V characteristics of the AlGaN/GaN HEMT measured at 1 MHz. The capacitance increases sharply near $V_{GS} \approx$ –1.5 V, indicating the threshold region where the 2DEG begins to form at the AlGaN/GaN interface. This transition point allows for the estimation of the threshold voltage, which was extracted to be approximately –1.5 V. The curve also helps determine gate capacitance values used later in mobility calculations.

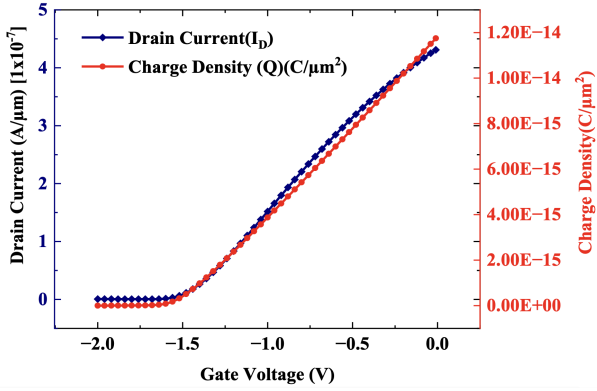

Fig. 3. Charge density and drain current vs. gate voltage for the AlGaN/GaN HEMT.

Fig. 3 shows the variation of drain current ($I_D$) and charge density (Q) with respect to gate voltage for the AlGaN/GaN HEMT. As the gate voltage becomes less negative, both $I_D$ and Q increase, indicating the formation of the two-dimensional electron gas (2DEG) at the AlGaN/GaN interface. The close correlation between charge density and current confirms that the gate effectively modulates carrier concentration. This plot reinforces the threshold voltage estimate (–1.5 V) and provides additional evidence of strong gate control over the channel. Such analysis also supports the accuracy of extracted parameters from C–V and transfer characteristics.

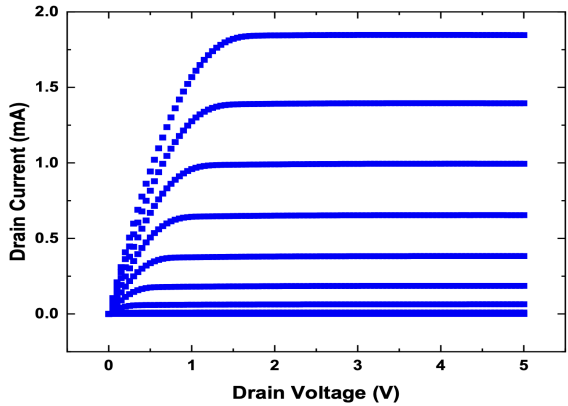

Fig. 4. Output characteristics ($I_D$–$V_d$) of AlGaN/GaN HEMT for varying gate voltages.

Fig. 4 presents the output characteristics ($I_D$–$V_{DS}$) of the AlGaN/GaN HEMT measured at various gate-source voltages ($V_{GS}$ from –2 V to 0 V, step size of 0.2 V). The plot shows typical HEMT behavior, with a linear region at low drain voltage followed by current saturation at higher $V_{DS}$. From this graph, the drain saturation voltage was observed to be approximately 1 V, which indicates the onset of velocity saturation and channel pinch-off. The maximum drain current ($I_{ON}$), extracted from the highest $V_{GS}$ curve, is 1.9 mA, reflecting the device's current-driving capability. The ON-resistance ($R_{ON}$) was calculated in the linear region using the expression:

$$R_{ON} = \frac{V_{DS}}{I_D}$$

Based on this relation, the extracted $R_{ON}$ was found to be approximately 22.72 Ω·µm, indicating low conduction loss and good efficiency for power applications.

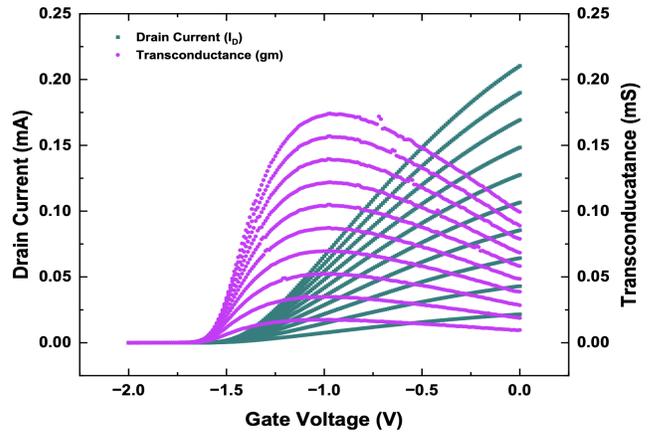

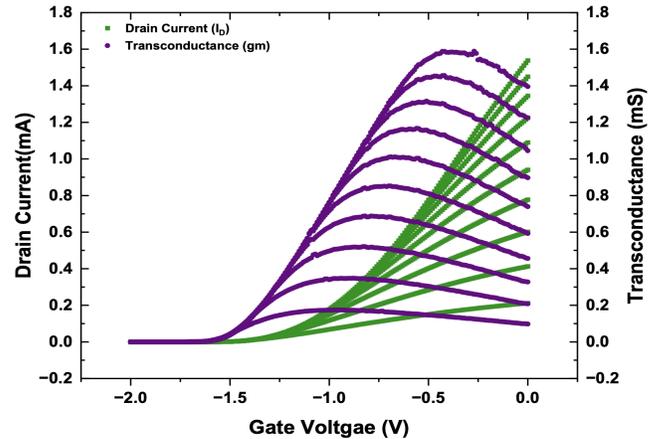

Fig. 5(a). Transfer characteristics and transconductance of AlGaN/GaN HEMT in the linear region, 5(b). Transfer characteristics and transconductance of AlGaN/GaN HEMT in the saturation region.

In Fig. 5(a), the drain current was measured with $V_{DS}$ swept from 10 mV to 100 mV, capturing the device behavior in the linear region, ideal for mobility extraction. The peak transconductance ($g_{m,max}$) in this region was extracted as 0.18 mS, which, combined with gate capacitance and device dimensions, is used to compute the field-effect mobility. Transconductance was calculated using the expression:

$$g_m = \frac{dI_D}{dV_{GS}}$$

This defines the slope of the transfer curve and quantifies how effectively the gate controls the channel current.

Fig. 5(b) presents the same characteristics in the saturation region, where $V_{DS}$ ranges from 100 mV to 1 V. The curves show stronger current conduction with a maximum transconductance of approximately 0.5 mS, indicating enhanced gain under higher drain bias. From these graphs, additional metrics such as threshold voltage, subthreshold swing (SS), and $I_{ON}/I_{OFF}$ ratio are extracted. The subthreshold swing was calculated using the relation:

$$SS = \left(\frac{d\,log_{10}I_D}{dV_{GS}}\right)^{-1}$$

This expression reflects how effectively the gate voltage modulates the drain current in the subthreshold region. The extracted SS value of 80 mV/decade indicates strong gate control and low leakage behavior.

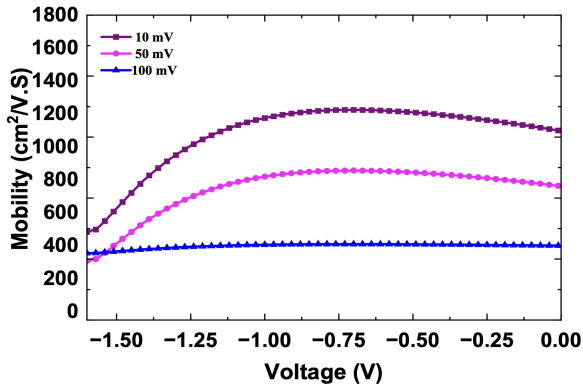

Fig. 6. Extracted field-effect mobility (μe) vs. gate voltage at different drain voltages (10 mV, 50 mV, 100 mV).

Fig. 6 presents the extracted field-effect mobility (μe) of the AlGaN/GaN HEMT using the transconductance method at drain voltages of 10 mV, 50 mV, and 100 mV. The mobility was calculated using the peak transconductance values from the linear region of the transfer characteristics, along with the estimated gate capacitance and device geometry. It was computed using the expression:

$$\mu_e = \frac{I_D/W}{Q*V_{DS}/L}$$

where $I_D$ is the drain current, L and W are the gate length and width, Q is the channel charge, and $V_{DS}$ is the drain-source voltage. The highest mobility value of approximately 1200 cm²/V·s was observed at $V_{DS}$ = 10 mV, and it decreased progressively at higher drain voltages. This trend highlights the strong dependence of carrier mobility on the applied drain bias.

At low drain voltages, the electric field across the channel remains weak, minimizing scattering effects and allowing carriers to move more freely. However, as $V_{DS}$ increases, the longitudinal electric field inside the channel strengthens, pushing carriers closer to the AlGaN/GaN interface. This results in enhanced interface roughness scattering, phonon scattering, and eventually velocity saturation, where carriers can no longer gain speed despite the increased electric field. Additionally, the rise in sheet carrier concentration further aggravates scattering mechanisms, particularly at higher fields where Coulombic interactions become more significant. These scattering events reduce the effective mobility and limit the overall current drive capability of the device. Furthermore, higher electric fields can lead to localized heating and non-uniform carrier transport, which further distorts mobility measurements. Therefore, accurate mobility extraction must be performed at low $V_{DS}$, where drift current dominates and carriers experience minimal perturbation. This ensures a more reliable estimation of the intrinsic material properties. These observations underscore the importance of carefully selecting bias conditions during measurement and the need for structural optimization such as barrier engineering or interface passivation to preserve high mobility and maximize GaN HEMT performance in real-world applications.

## IV. CONCLUSIONS

This study presented a combined simulation and experimental analysis of GaN HEMT to evaluate their electrical performance, with a particular focus on mobility extraction. Energy band simulations confirmed the formation of a two-dimensional electron gas (2DEG), while C–V measurements identified a threshold voltage near –1.5 V. Transfer characteristics in the linear region enabled the extraction of peak transconductance (0.18 mS), which was used to compute the field-effect mobility using the transconductance method. The maximum extracted mobility was approximately 1200 cm²/V·s at low drain bias, highlighting the device's potential for high-speed and high-efficiency operation. Output characteristics revealed an ON current of 1.9 mA and a saturation voltage of 1 V. Additional parameters, including subthreshold swing (80 mV/decade), DIBL (5 mV/V), and ON-resistance (22.72 Ω·µm), further confirmed the device's suitability for high-power and RF applications. Overall, the integration of simulation and measurement provided a comprehensive understanding of the structure-performance relationship, with mobility extraction serving as a key indicator of device quality.

## ACKNOWLEDGMENT

The author sincerely thanks Dr. Taewoo Kim for generously providing experimental data used in this report. His support and contribution through access to his research team's work were instrumental in enabling mobility extraction analysis for GaN HEMT.The author is also grateful for his insightful guidance and encouragement throughout the project. Appreciation is also extended to the Department of Electrical and Computer Engineering at Texas Tech University for providing the necessary resources and support.

## REFERENCE

[1] Mishra, Umesh K., Primit Parikh, and Yi-Feng Wu. "AlGaN/GaN HEMTs-an overview of device operation and applications." Proceedings of the IEEE 90.6 (2002): 1022-1031.

[2] del Alamo, Jesús A., and Jungwoo Joh. "GaN HEMT reliability." Microelectronics reliability 49.9-11 (2009): 1200-1206.

[3] Trew, Robert J., Daniel S. Green, and Jeffrey B. Shealy. "AlGaN/GaN HFET reliability." IEEE Microwave magazine 10.4 (2009): 116-127.

[4] Kikkawa, Toshihide, et al. "High performance and high reliability AlGaN/GaN HEMTs." physica status solidi (a) 206.6 (2009): 1135-1144.

[5] Dammann, Michael, et al. "Reliability of AlGaN/GaN HEMTs under DC-and RF-operation." 2009 Reliability of Compound Semiconductors Digest (ROCS). IEEE, 2009.

[6] Inoue, Takashi, et al. "30-GHz-band over 5-W power performance of short-channel AlGaN/GaN heterojunction FETs." IEEE transactions on microwave theory and techniques 53.1 (2005): 74-80.

[7] Wu, Y-F., et al. "8-watt GaN HEMTs at millimeter-wave frequencies. "IEEE InternationalElectron Devices Meeting, 2005. IEDM Technical Digest.. IEEE, 2005.

[8] Lenka, T. R., and A. K. Panda. "Characteristics study of 2DEG transport properties of AlGaN/GaN and AlGaAs/GaAs-based HEMT." Semiconductors 45.5 (2011): 650-656.

[9] Osipov, Konstantin, et al. "Local 2DEG density control in heterostructures of piezoelectric materials and its application in GaN HEMT fabrication technology." IEEE Transactions on Electron Devices 65.8 (2018): 3176-3184.

[10] He, Jiaqi, et al. "Recent advances in GaN‑based power HEMT devices." Advanced electronic materials 7.4 (2021): 2001045.

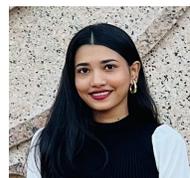

**Tanjim Rahman** is a Master's student in Electrical and Computer Engineering at Texas Tech University. She received her B.Tech in Electronics and Communication Engineering from NIT Silchar, India, in 2024. Her research focuses on TCAD simulation and characterization of GaN HEMTs and LEDs.